\documentclass[12pt]{article}
\usepackage{putex}
\usepackage{graphicx}

\def\beq{\begin{equation}} \def\eeq{\end{equation}}
\def\beqn{\begin{equation*}} \def\eeqn{\end{equation*}}
\def\bea{\begin{eqnarray}} \def\eea{\end{eqnarray}}
\def\nn{\nonumber}

\begin{document}


\thispagestyle{empty}

\begin{flushright}\scshape
March 2010
\end{flushright}
\vskip1cm

\begin{center}

{\LARGE\scshape Remarks on the consistency of minimal deviations from General Relativity
\par}
\vskip15mm

\textsc{Josep M. Pons$^{ac}$ and Pere Talavera$^{bc}$}
\par\bigskip
$^a${\em
Departament d'Estructura i Constituents de la Mat\`eria,
Universitat de Barcelona,\\
Diagonal 647, E-08028 Barcelona, Spain.}\\[.1cm]
$^b${\em
Departament de F{\'\i}sica i Enginyeria Nuclear,
Universitat Polit\`ecnica de Catalunya,\\
Comte Urgell 187, E-08036 Barcelona, Spain.}\\[.1cm]
$^c${\em
Institut de Ciencies del Cosmos,
Universitat de Barcelona,\\
Diagonal 647, E-08028 Barcelona, Spain.}\\[.1cm]
\vspace{5mm}
\end{center}

\section*{Abstract}

We study the consequences of the modification of the phase space structure of General Relativity imposed by breaking the full diffeomorphism invariance but retaining the time foliation preserving diffeomorphisms. We examine the different sectors in phase space that satisfy the new structure of constraints. For some sectors we find an infinite tower of constraints. In spite of that, we also show that these sectors allow for solutions, among them some well known families of black hole and cosmologies which fulfill all the constraints. We raise some physical concerns on the consequences of an absolute Galilean time, on the thermodynamical pathologies of such models and on their unusual vacuum structure.

\vspace{3mm} \vfill{ \hrule width 5.cm \vskip 2.mm {\small
\noindent E-mail: pons@ecm.ub.es, pere.talavera@icc.ub.edu }}

\newpage
\setcounter{page}{1}

\section{Motivation, outlook and conclusions}

There has been recently a considerably huge activity regarding different aspects of 
an eventually renormalizable gravitational theory  \cite{Horava:2009uw}. 
Despite the immense amount of work that the subject has triggered, only a few authors
have dealt in detail with the consistency of the initial proposal \cite{Charmousis:2009tc,Blas:2009yd, Iengo:2009ix,Henneaux:2009zb} while the vast majority of them deal with 
applications in cosmology \cite{Calcagni:2009ar,Brandenberger:2009yt} and with the obtention of some solutions, typically applicable to black hole physics \cite{Lu:2009em}.

It is our aim to tidy up some dangling issues and fill technical details of the constrained system originated from the simplest version contained in the spirit of the initial proposal \cite{Horava:2009uw}. Based on those findings we will construct a sample of explicit cosmological vacuum models consistent with the new dynamics.  Our final goal is to show in a cristal clear way that giving up full diffeomorphism invariance, in a very specific way, implies some bizarre and so far unexplored consequences at the most fundamental level.

We will always compare our outputs with the {\sl deep} infrared theory we deal with, General Relativity (GR) -with cosmological constant- \cite{Arnowitt:1962hi},
\beq   
{\cal L}_{\!{}_{GR}} = \sqrt{\gamma} N \Big({}^{{}^{(3)}\!}\!R + {K}_{ij}{K}^{ij}-{K}^2 - 2 \Lambda\Big)\,,
\label{thelagGR}
\eeq
where there is full $4$-diffeomorphism invariance $x^\mu\to x^\mu-\epsilon^\mu$, with $\epsilon^\mu$ an arbitrary infinitesimal function of the coordinates as well as of the fields.
The simplest setting one can find in the literature that mimics the partial breaking of the symmetry of time diffeomorphisms is just a {\sl small} modification in the kinetic term of (\ref{thelagGR}), see section 2.
Notice however that even such a {\it slight} modification dramatically modifies the constraint's structure of the theory in phase space and makes it inconsistent in most classical settings of the initial conditions, with the sure exception of some symmetric, protected cases and of course by dimensionally reducing it to a one dimensional mechanical model; this shows how {\it finely tuned} is GR for it to be dynamically consistent. 

A general consequence of the modification of (\ref{thelagGR}) - or the more general modifications discussed in the literature - is that, due to the reduction of diffeomorphism invariance, the foliation of the spacetime in equal-time surfaces is fixed and acquires a direct physical meaning: there is an absolute concept of simultaneity of events. In this sense the time in these modified theories of gravity is a Galilean time.

As other consequences of this partial breaking of diffeomorphism invariance we may list some of our findings, see section \ref{swsol}
\begin{enumerate} 
\item Due to the impossibility to perform time diffeomorphism with spatial dependences, one can not construct a Rindler causal horizon through each spacetime point and hence nor a Unruh radiation can be defined.
\item Whereas the de Sitter solution in pure GR with cosmological constant accepts several types of time foliations (open, closed and flat $3$-slices) but the physics remains always the same, we find that in the modified theory of gravity, every type of foliation corresponds to a different physical vacuum. Hence the vacuum is degenerate.
\item Some specific solutions of GR are easily seen to be solutions of the new dynamics as well, but one should be aware that this is not a coordinate independent statement.
We will give examples of background configurations of GR that in some coordinate incarnation are indeed solutions of the minimally modified dynamics, whereas they are no longer so when written in other systems of coordinates.
\item Contrary to some remarks in the literature, the implementation of the partial gauge fixing of the projectability conditions -i.e., that the lapse only depends on the time coordinate -, has no effect whatsoever on the Hamiltonian constraint, if one sticks to the rules of deriving the dynamics from the action principle.
\end{enumerate}
Due to a certain confusion in the literature, let us make an additional comment on the last remark above. One can interpret the projectability condition in two different ways. The natural way, which we advocate here, is to understand it as a partial gauge fixing for a theory defined by an action, with no preconceptions on the possible coordinate dependences of the fields. We may call it the {\sl soft} projectability condition. Another view, the {\sl hard} projectability condition, assumes a modification of the original action, in that it already contains the restriction that the lapse only depends on time. We believe that this second point of view, though legitimate, is less natural than the first one. We will be back to this issue in subsections \ref{lookfortert} and \ref{projec}.

Our analysis will be roughly split into two main venues, see section \ref{sta}:\\ \noindent
{\sl i)} In the generic case the constraint analysis can be performed to its end, and one obtains as a result a tertiary constraint and a partial determination of the Lagrange multiplier $\lambda^0$ in the Dirac Hamiltonian. But one must be aware that other retrictions, coming form boundary conditions, may produce a collapse of the dynamics, in the sense for instance of the compulsory vanishing of the lapse \cite{Henneaux:2009zb}. \\
\noindent
{\sl ii)} In non generic cases, we show that the typical situation is that of an indefinite chain of tertiary, quartiary, etc, constraints which seem to set notable limitations to the allowed initial conditions. Even if we show some settings that allow for solutions, it is quite likely that there is inconsistency for a large class of initial conditions, much beyond the situation in GR.

In addition of the mentioned bizarre consequences derived from partial diffeomorphism breaking and the 
increasing experimental evidence on the correctness of GR in the deep infra-red \cite{reyes} it is fair to mention that although in principle appealing, a weakness of the proposal is that there seems to be no
clear mechanism guarantying that the low energy limit of this modified theory is going to restore full fledged GR.

\section{Set up: minimal deviation from General Relativity} 
\label{setup}

 In order to make our point more clear we will follow the approach of \cite{Sotiriou:2009bx} which in some 
sense is based in some sort of {\sl effective} theory construction. Such approach is theoretically robust and 
it is suitable from the phenomenological point of view.
In doing so we will encounter two problems that we will refer below. The construction goes as follows: one introduces an anisotropic scaling at the ultraviolet fixed point from which one constructs a 
power-counting renormalizable theory organized in terms of a placeholder symbol $\kappa$, which allows for a kind of {\sl weighted scaling dimensional analysis}, 
\begin{eqnarray}
[dx]= [\kappa]^{-1}\,,\quad  [dt]= [\kappa]^{-3}\,,\quad [N]=[\kappa]^{0} \,,\quad [N^i]=[\kappa]^2 \,,\quad [\gamma_{ij}]=[\kappa]^{0}\,.
\end{eqnarray}

At high energy this scaling modifies the kinetic part of (\ref{thelagGR}) with the presence of a new parameter and simultaneously introduce a plethora of new relevant operators in the potential term. Most of these terms become irrelevant in the infrared. In most of applications one assumes that a {\sl naive} power counting in derivatives holds and then one can discard the contribution of the tower of operators present in the potential in the infrared. Unfortunately their coefficients have a logarithmic running spoiling this behavior unless its ultraviolet value is unnaturally small \cite{Iengo:2009ix}, this is the first of the problem referred above. The second one, already stressed in \cite{Charmousis:2009tc}, is that the inclusion of the Lorentz violation term in the kinetic part of the Lagrangian leads to an extra scalar mode for the graviton at all momenta and thus if one does not make corrections for these extra degrees of freedom one will never recover General Relativity (GR) in the infrared. A proposal has been made to amend such behavior \cite{Blas:2009qj} but some criticism on the results has been raised \cite{Mukohyama:2009tp}.

To avoid the first of these problems we will only focus in the possible deviations of GR near the infrared, thus in some sense we will give up issues concerning the ultraviolet completion of the theory. With respect to the second problem we will not be concerned with it and explore only a {\sl minimal deviation} from GR in the spirit of the original proposal \cite{Horava:2009uw}. Given the previous line of reasoning, the starting point (\ref{thelag}), see below, is the simplest of the settings discussed in \cite{Sotiriou:2009bx} and corresponds to a set up where only relevant, dimension four operators in the infra-red are kept. The kinetic term will contain a {\sl slight} difference with respect to GR while the potential term will remain the same.

Let us remind of the  4-dimensional metric decomposition in terms of the 3-dimensional metric and the lapse and shift fields
\beq g_{\mu\nu}= \left(\begin{array}{c|c}- N^2+N^i N_{i}  & N_{i} \\ \hline N_{i} & \gamma_{ij}  
\end{array}\right)\,,
\eeq
which implies $\sqrt{-g}= \sqrt{\gamma}N$.
Spatial indices are raised and lowered with $\gamma_{ij}$ and its inverse $\gamma^{ij}$ respectively . 
The Lagrangian takes the form
\beq   
{\cal L} = \sqrt{\gamma} N \Big({}^{{}^{(3)}\!}\!R + {K}_{ij}{K}^{ij}-\lambda{K}^2- 2 \Lambda\Big)\,,
\label{thelag}
\eeq
where ${}^{{}^{(3)}\!}\!R$ is the  scalar curvature for the $3$-metric $\gamma_{ij}$,  written henceforth simply as $R$, and ${K}_{ij}$ is the extrinsic curvature of the equal time surfaces,
\beq 
K_{ij}={1\over 2N}\left(\dot \gamma_{ij}-\nabla_i N_j-\nabla_j N_i\right)\,,
\label{Kij}
\eeq
with $K$ been its trace, $K= {K}^{ij}\gamma_{ij}$. 
 Notice that for the value $\lambda =1$ we recover the standard ADM Lagrangian  (\ref{thelagGR}) with 
all its virtues and properties. Thus the initial {\sl assumption} in \cite{Horava:2009uw}  is that as one goes to the infrared and recovers full diffeomorphism invariance the parameter $\lambda$ must run exactly to one.  For $\lambda\neq 1$ the invariance is reduced to the sugbroup of foliation preserving diffeomorphisms, for which the  time diffeomorphism variation $\delta x^0=:-\epsilon^0$ only depends on the time coordinate $x^0$. From the phenomenological point of view the value of $\lambda$ is restricted to lie very near the GR value \cite{Dutta:2010jh}. In this sense we will take $\lambda$ as a varying parameter that {\sl only} in the deep infrared matches the GR value. For the sake of completeness we will explore also the region where $\lambda$ differs significanlty form $1$.

In order to obtain the Hamiltonian dynamics we define first the variables in phase space.
Since the Lagrangian (\ref{thelag}) does not depend on the time derivatives of the lapse and shift, we identify the momenta $P_\mu$ conjugate to the lapse $N=:N^0$ and shift $N^i$ as primary constraints in phase space
\beq
P_\mu\simeq 0\,.
\eeq
The Lagrangian definition of the momenta $\pi^{ij}$ conjugate to $\gamma_{ij}$ gives
\beq
\pi^{ij} = \frac{\partial{\cal L} }{\partial \dot\gamma^{ij}} = \sqrt{\gamma}\Big({K}^{ij} - \lambda  K \gamma^{ij}   \Big)\,.
\label{pidef}
\eeq
Note that $\pi^{ij}$ is a tensor density of weight $1$ with respect to $3$-diffeomorphisms. To prepare for the construction of the canonical Hamiltonian, we trade the canonical variables $\pi^{ij}$ for 
${K}^{ij}$. The trace of (\ref{pidef}) is ($\pi:= \pi^{ij}\gamma_{ij}$)
\beq
\pi =  \sqrt{\gamma}(1-3\lambda) K\,.
\label{pitrace}
\eeq
At this point we notice that a special behavior takes place at $\lambda=\frac{1}{3}$. In this case a new primary constraint appears, $\pi \simeq 0$. What happens is that, for this value of $\lambda$, the gauge symmetry of the theory is enhanced: this is the anisotropic Weyl invariance. It is obvious that to a new gauge invariance there should be associated new first class constraints, as the eventual generators. We will leave aside momentarily this particular case and come back to it in section \ref{crit}.

Using (\ref{pidef}) and (\ref{pitrace}) we can isolate ${K}^{ij}$ in terms of the canonical variables,
\beq
{K}^{ij} = \frac{1}{\sqrt{\gamma}}(\pi^{ij} +  \frac{\lambda}{1-3\lambda} \pi \gamma^{ij})\,.
\eeq

With all the previous inputs, we can rewrite the system (\ref{thelag}) in the Hamiltonian formalism. This will allow us to handle the stabilization - i.e. time preservation - of the constraints in phase space.
The canonical Hamiltonian is defined by the spatial integration $H= \int d{\bf x}\, {\cal H}$, with 
${\cal H} = \pi^{ij}\dot \gamma_{ij} - {\cal L}$. We obtain, up to boundary terms
\beq
{\cal H} =  \sqrt{\gamma}N\Big(-R  +  \gamma^{-1} \pi^{ij}\pi_{ij} +  \frac{\lambda}{1-3\lambda}  \gamma^{-1}  \pi^2   + 2 \Lambda  \Big) - 2 N_i\, \nabla_j\pi^{ij} \,.
\label{realham}
\eeq
Defining $\beta:=  \frac{1-\lambda}{2(1-3\lambda)}$, the Hamiltonian density can be cast as
\beq
{\cal H} ={\cal H}^{\!{}^{(ADM)}} +\beta \frac{N}{\sqrt{\gamma}}\pi^2\,,
\label{hamilden}
\eeq
which for $\beta=0$ reduces to the standard ADM formulation including the cosmological constant term.

\subsection{The secondary constraints }

The secondary constraints are obtained under the requirement that the primary ones, $P_\mu$, are preserved under the time evolution. By varying the action with respect to the lapse we obtain the Hamiltonian constraint
\beq
{\cal H}_0 =  \sqrt{\gamma}\Big(-R  +  \gamma^{-1} \pi^{ij}\pi_{ij} +  \frac{\lambda}{1-3\lambda}  \gamma^{-1}  \pi^2   + 2 \Lambda  \Big) \,,
\label{hamconst}
\eeq
while the variation with respect to the shift leads to the momentum constraints
\beq
{\cal H}_j =  - 2 \nabla_i\,\pi^{i}_{\ j} \,.
\label{momconst}
\eeq
Notice that the only difference with respect to the ADM formulation lies solely in ${\cal H}_0$.  All in all we have
\beq
{\cal H}_0 ={\cal H}^{\!{}^{(ADM)}}_0 +\beta \frac{\pi^2}{\sqrt{\gamma}},
\qquad {\cal H}_j={\cal H}_j^{\!{}^{(ADM)}}\,,
\label{ADMcomp}
\eeq
where $\beta$ parametrizes the deviation with respect to GR.
The Dirac Hamiltonian can be expressed as
\beq
H_D = \int d{\bf x} (N^\mu {\cal H}_\mu +\lambda^\mu P_\mu)\,,
\label{dirham}
\eeq
with the Lagrange multipliers $\lambda^\mu$ being - as of now - arbitrary functions.

\subsection{ The tertiary constraint }
\label{lookfortert}

Now we must look for tertiary constraints, as the consequence of the  stabilization of the secondary ones (\ref{ADMcomp}), ${\cal H}_\mu$. This computation has been properly addressed in \cite{Blas:2009yd} within the Lagrangian framework in the tangent bundle\footnote{Actually, from the Lagrangian point of view this will be a secondary constraint, because a $n$-ary constraint in phase space has a corresponding $(n-1)$-ary constraint - through the pullback operation - in the tangent bundle. Primary constraints in phase space correspond to identities in the tangent bundle.}.
Instead, we will reobtain the tertiary constraint working within the canonical formalism. 

The stabilization of the secondary constraints takes the form of the requirement 
\beq
\left\{{\cal H}_\mu,\, H_D\right\} = \left\{{\cal H}_\mu,\, \int d{\bf x}' N^\nu {\cal H}_\nu({\bf x}')\right\}\simeq 0\,.
\label{secest}
\eeq
Using (\ref{dirham}) we can display the different terms that contribute to (\ref{secest}). Two relevant pieces of information help to calculate (\ref{secest}). They are: {\sl i)} Notice first of all that due to the standard behaviour of a scalar density under $3$-diffeomorphisms one has
\beq
\left\{{\cal H}_0,\,\int d{\bf x}' N^i{\cal H}_i({\bf x}')\right\} = \partial_i(N^i{\cal H}_0)\,,
\eeq 
which vanishes in the primary and secondary constraints' surface.  {\sl ii)} On the other hand, the algebra of the generators of $3$-diffeomorphisms closes in the standard way because they coincide with those of the ADM case. This in turn guaranties that the stabilization of the constraints ${\cal H}_i$ does not introduce new constraints.  Using these two facts the only relevant term in the computation becomes
\beq 
{\rm tertiary\ constraint\ } =\left\{{\cal H}_0({\bf x}),\,\int d{\bf x}' N{\cal H}_0({\bf x}')\right\}\,.
\label{stab}
\eeq

The computation of (\ref{stab}) is facilitated by the smearing of the constraint ${\cal H}_0$ by means of an arbitrary function $\eta$, which we take of compact support.
So our task is reduced to the computation of 
\beq 
\int d{\bf x}\,\eta\,({\rm tertiary\ c.}) =\left\{\int d{\bf x}\,\eta\,{\cal H}_0({\bf x}),\,\int d{\bf x}' N{\cal H}_0({\bf x}')\right\}\,.
\label{tc}
\eeq
 Using (\ref{ADMcomp}), we expand the rhs of (\ref{tc}) as 
 \begin{equation}
 \int d{\bf x}\,\eta\, 
({\rm tertiary\ c.\ })={\cal O}(\beta^0)+{\cal O}(\beta)+{\cal O}(\beta^2)\,.
\end{equation}
The ${\cal O}(\beta^0)$ term in the above expression is proportional to the momentum constraints and plays no role. The ${\cal O}(\beta^2)$ term vanishes because no derivatives of the fields are involved. Thus the tertiary constraint stems only from the terms ${\cal O}(\beta)$.
\bea
\int d{\bf x}\,\eta\,({\rm tertiary\ c.})
= \beta \left( \left\{\int d{\bf x}\,\eta \frac{\pi^2}{\sqrt{\gamma}},\,\int d{\bf x}' N{\cal H}^{\!{}^{(ADM)}}_0 \right\} -\left\{\int d{\bf x} N \frac{\pi^2}{\sqrt{\gamma}},\,\int d{\bf x}'\,\eta{\cal H}^{\!{}^{(ADM)}}_0 \right\}\right)\,.
\label{tert}
\eea
Consider the first term  in the r.h.s of (\ref{tert}).  Getting rid momentarily of the smearing, we realize that it  corresponds to the time derivative of 
$\frac{\pi^2}{\sqrt{\gamma}}$ under the evolution provided by the ADM Hamiltonian in the gauge $N^i=0$
\begin{equation}
\left\{ \frac{\pi^2}{\sqrt{\gamma}},\,\int d{\bf x}' N{\cal H}^{\!{}^{(ADM)}}_0 \right\}\,.
\label{timeder}
\end{equation}
Thus we can compute it with standard formulas - see for instance \cite{Arnowitt:1962hi,Wald:1984rg} - for the time derivative of the components of the $3$-metric and their conjugate momenta. We notice then  
the crucial fact that, when smearing with $\eta$, all terms in (\ref{timeder}) that have no 
derivatives of the fields will cancel against the second contribution to (\ref{tert}). It turns out that
\beq
\Big(\frac{\pi^2}{\sqrt{\gamma}}\Big)^{\centerdot} = {\rm irrelevant\ terms\ } - 4 \pi \triangle N \,,
\label{oldodts}
\eeq
where $\triangle N$ stands for the Laplacian $\gamma_{ij}\nabla^i\nabla^j N$. 
 The relevant contribution in (\ref{oldodts}) arises  entirely from the $\dot\pi$ term. 
All in all,  the contribution to (\ref{tert}) becomes
\bea
\int d{\bf x}\,\eta\,({\rm tertiary\ c.})
= - 4\, \beta \int d{\bf x}\Big(\eta \pi  \triangle N - N  \pi  \triangle  \eta \Big)\,
= 4\, \beta  \int d{\bf x}\eta\Big(2 \nabla_i \pi \partial^i N + N  
\triangle\pi\Big)\,,
\label{2view}
\eea
where in the last equality we have used part integration and the fact that the function $\eta$ is of compact support. From the above expression the tertiary constraint is identified as
\beq
2 \nabla_{\!i} \pi\, \partial^i N + N  \triangle\pi\simeq 0\,.
\label{thetert}
\eeq
This equation, (\ref{thetert}), has been first obtained in \cite{Kocharyan:2009te}. It has also been considered in \cite{Henneaux:2009zb}, although not really interpreted as a constraint because the momenta $P_\mu$ canonically conjugate to $N^\mu$ are eliminated in \cite{Henneaux:2009zb} and the lapse and shift variables take over the role of Lagrange multipliers. We prefer for now to keep all the variables of the formalism and the Hamiltonian (\ref{dirham}). 

At this point it is worth noticing that from the perspective of the {\sl hard} projectability condition, there is no tertiary constraint at all, because in such case one must stabilize  $\int d{\bf x}\,{\cal H}_0$, which is like redoing (\ref{2view}) with $\eta = 1$; then the computation in the first equality of  (\ref{2view}) gives a vanishing result: $\eta \pi  \triangle N - N  \pi  \triangle  \eta =0$. In conclusion: there is no tertiary constraint for the {\sl hard} projectability condition implemented in the Lagrangian (\ref{thelag}).
\section{Stabilization of the tertiary constraint }
\label{sta}

The stabilization of (\ref{thetert}) under the time evolution will produce the appearance of $\dot N$, which, according to the dynamics, is the multiplier $\lambda^0$ in the Dirac Hamiltonian (\ref{dirham}). Thus the stabilization of (\ref{thetert}) has the form
\beq
\triangle\pi\, \lambda^0 + 2 \nabla_i \pi\, \partial^i\lambda^0 + {\rm (terms\ with\ no\ \lambda^0)} =0\,,
\label{lameq}
\eeq
where we have taken the equality to zero as an ordinary equality because it is legitimate to make 
the determination of the multipliers $\lambda^\mu$ on shell. Note that there are several possible choices in the space of field configurations that lead  
to the fulfillment of (\ref{thetert}) and (\ref{lameq}). For instance, if we consider a configuration satisfying $\nabla_{\!i} \pi=0$, then (\ref{thetert}) holds and (\ref{lameq}) is independent of the value of $\lambda^0$, but not void.

In this section we will consider different consistent ways for which the tertiary constraint is enforced by restricting the field configurations in a way that guaranties (\ref{thetert}).

\subsection{The generic case: $\triangle\pi\neq 0 $}
\label{caseno}

As a first trial let us consider a generic case with $\triangle\pi\neq 0$. Notice that if  {\sl i)} the initial conditions for the fields at, say, $t=0$, satisfy all the constraints, including (\ref{thetert}), and if  {\sl ii)} the multiplier $\lambda^0$ in the Dirac Hamiltonian (\ref{dirham}) is chosen so that it satisfies (\ref{lameq}), then it is guarateed that all constraints will be satisfied at any time. Which means that the constraint analysis is finished. Notice though that the analysis made here is based on local requirements. What may remain to be explored is the adequacy of these results for certain boundary conditions. Some concerns in this respect will be drawn below.

In this generic case one can obtain, at least formally, the partial determination of the multiplier $\lambda^0$ as follows:
First factorize out ${\triangle\pi}$ from (\ref{lameq}) and obtain an expression of the form
\beq
\lambda^0 = {\vec V} \lambda^0 + U\,,
\label{lameq2}
\eeq
with ${\vec V}$ the differential operator $\displaystyle {\vec V} = - 2 \frac{\nabla_i \pi}{\triangle\pi} 
\partial_i$ and $U$ the remaining quantity. 
The solution to (\ref{lameq2}) will be the sum of a particular solution plus an arbitrary solution of the associated homogeneous equation. 
As long as the vector field ${\vec V}$ is diferent form zero - which must be, because we are in a generic configuration and so ${\vec\nabla \pi}\neq 0$ -, one can locally change to a spatial coordinates system such that ${\vec V}=\partial_z$. 
Then the general solution of (\ref{lameq2}) is
\begin{equation}
\lambda^0 (x,y,z,t)= - \int^z dz'  e^{z-z'} U((x,y,z',t)) + e^z f(x,y,t)\,.
\end{equation}
 
As already mentioned, our determination of $\lambda^0$ is {\sl only} of  
local validity and one may ask on the global behaviour of the function $\lambda^0$, which is an issue not addressed here and that essentially depend on the specifics of the boundary conditions that are imposed. This behaviour will affect that of $N$, since $\dot N=\lambda^0$. Within the very same generic case and assuming an asymptotically flat space the behaviour of $N$ has been studied directly from the analysis of (\ref{thetert}) in \cite{Henneaux:2009zb}. The outcome of the analysis is that to prevent the lapse function from been divergent at the boundary, the only acceptable solution is $N=0$. 

The cases to be considered in the following subsections are non generic, that is, they satisfy $\triangle\pi= 0$. 

\subsection{Preserving the first class condition.}
\label{newimpl}

Let us consider a second trial set for the fulfillment of the tertiarty constraint (\ref{thetert}). Since the consistency of the whole picture requires that $N\neq 0$, otherwise the Lagrangian vanishes and the dynamics disappears, we can take the tertiary constraint in the form
\beq
\Phi := 2 \nabla_{\!i} \pi \,\frac{\partial^i N}{N} +   \triangle\pi\simeq 0\,,
\label{thetert2}
\eeq
which is more suited for our present purposes. Note that except for some special circumstances, this 
constraint will make the primary constraint $P_0:=P$ second class,
\beq
\{P({\bf x}),\,\Phi({\bf y})\} = \frac{\nabla^i \pi}{N}({\bf y})\Big(\frac{\partial_i N}{N}({\bf 
y})\delta({\bf x-y}) - 2 \nabla_{\!i}^{{}^y}\delta({\bf x-y})\Big)\ne 0\,,
\eeq
even in the case when the projectability condition holds\footnote{Notice however that being second 
class in field theory is more tricky than in the mechanical case. Indeed, if we were in mechanics, a 
second class condition for $P$ will undoubtedly determine the multiplier $\lambda^0$. As we have just seen, this is not the case in field theory.} unless $\nabla^i \pi$ vanishes on shell. This last situation amounts to impose the new set of constraints
\begin{equation}
\label{pcond}
\nabla_{\!i} \pi\simeq 0\,,
\end{equation}  
which already imply (\ref{thetert2}). As a matter of fact, there is another reason 
for analyzing this case: if we were to require that (\ref{thetert}) introduces no restriction upon the 
lapse, then we should impose (\ref{pcond}). Since the partial spatial derivatives commute with the dynamics, it is more convenient to use the form 
\beq
\partial_i(\frac{\pi}{\sqrt{\gamma}}) \simeq 0\,,
\label{thetert3}
\eeq
in order to explore the stabilization of (\ref{pcond}).
After using the Hamiltonian constraint to eliminate terms with $\pi^{kl}\pi_{kl}$, the new constraints originated from the stabilization of (\ref{thetert3}) become
\beq
\partial_i\left[ N\left(2 R - \frac{1}{1-3\lambda}\,\frac{\pi^2}{\gamma}- 6 \Lambda\right) -  2\triangle N  \right]\simeq 0\,,
\label{thefourth}
\eeq
or, since $\displaystyle\partial_i(\frac{\pi^2}{\gamma})$ vanishes on shell by virtue of (\ref{thetert3}),
\beq
\frac{\partial_i N}{N} \left( R - \frac{1}{2(1-3\lambda)}\,\frac{\pi^2}{\gamma}- 3 \Lambda\right) +   \partial_i R
- \frac{\partial_i \triangle N}{N}  \simeq 0\,,
\label{thefourthdos}
\eeq
thus showing that the stabilization of (\ref{pcond}) has introduce a new constraint (\ref{thefourthdos}). Note that the initial goal to prevent the constraint $P_0$ to become second class faces new challenges. It seems that 
the most simple setting aimed at this contains as an ingredient the projectability condition, i.e. the lapse only depends on the time coordinate,
together with the requirement that the $3$-surfaces - labeled by the time coordinate - of the foliation of the spacetime must be surfaces of constant curvature, that is,
\beq
\partial_i N\simeq 0\,,\qquad \partial_i R\simeq 0\,.
\label{thefourth3}
\eeq
But we are not over yet, 
because, again, stability must be required to these constraints. In this new analyisis we must consider 
{\sl i)} The stabilization of the projectability condition in (\ref{thefourth3}) results in the multiplier 
$\lambda^0$ satisfying $\partial_i\lambda^0 =0$ on shell, which is just fine because the only time-diffeomorphisms permitted are those of the type $x^\mu\to x^\mu-\delta^\mu_0\epsilon^0(x^0)$. {\sl ii)}  As regards the stability of $\partial_i R\simeq 0$, notice 
that since $R$ is a $3$-scalar, we know that $\{R,\, \int N^i {\cal H}_i \} = N^i\partial_i R$, which 
already vanishes on shell by virtue of (\ref{thefourth3}). Thus we only need to compute $\{\partial_i R,\, \int N {\cal 
H}_0 \}$. This results in the new set of constraints
\beq
\partial_i (\frac{R^{kl}\pi_{kl}}{\sqrt{\gamma}})\simeq 0\,.
\label{thefifth}
\eeq
Other constraints will follow from the stabilization of (\ref{thefifth}). We will not pursue this route any 
further, but make a general comment. There seems to be no obstacle for the stabilization mechanism originated from (\ref{thetert3}) to go on indefinitely, giving at every stage new constraints of the 
form of the gradient of a scalar. Notice that this scalar will involve more derivatives the more we advance in the process.

\subsection{The projectability condition}
\label{projec}

The third and last scenario we will focus on for the fullfilment of the tertiary constraint (\ref{thetert}) assumes that the projectability condition is satisfied. In this case, as long are we keep $N\neq 0$, a new constraint is compulsory, namely, 
\beq
\triangle\pi \simeq 0\,.
\label{triangpi}
\eeq
The stabilization of the projectability condition goes along the very same lines as in the previous case, subsection \ref{newimpl}.  As regards (\ref{triangpi}), the easiest way to satisfy it is to asume the stronger condition (\ref{thetert3}), in which case we will end up with the scenario discussed in \ref{newimpl}. Going back to the stabilization of (\ref{triangpi}), and following the same argument as with (\ref{thetert3}), we will stabilize
\beq
\sqrt{\gamma}\triangle (\frac{\pi}{\sqrt{\gamma}})\simeq 0\,.
\label{triangpi1}
\eeq
Since $\triangle(\frac{\pi}{\sqrt{\gamma}})$ is a $3$-scalar, we already know that
$\{\sqrt{\gamma} \triangle (\frac{\pi}{\sqrt{\gamma}}),\,\int N^i {\cal H}_i \} = \partial_i\Big(N^i\triangle (\frac{\pi}{\sqrt{\gamma}})\Big) \simeq 0$, and we must only compute $\{ \sqrt{\gamma}\triangle (\frac{\pi}{\sqrt{\gamma}}),\,\int N {\cal H}_0 \}$. Keeping $N\neq 0$ the final result is 
\beq
\Big(\sqrt{\gamma}\triangle (\frac{\pi}{\sqrt{\gamma}})\Big)^{\centerdot} \simeq 
2 N \Big( \sqrt{\gamma} \triangle R - \partial_i(\pi^{ij}\partial_j(\frac{\pi}{\sqrt{\gamma}}))\Big)
\,,
\label{projectablecase2}
\eeq
and thus the new constraint, byproduct of the stabilization of (\ref{triangpi1}), is 
\beq
 \sqrt{\gamma} \triangle R - \partial_i(\pi^{ij}\partial_j(\frac{\pi}{\sqrt{\gamma}}))\simeq 0\,.
\label{projectablecase3}
\eeq
As in the previous case the stabilization mechanism goes on indefinitely, and an infinite tower of new constraints, with more and more derivatives involved, appears as we advance in the algorithmic procedure. 

In some versions of the original model one implements the projectability condition from the very begining. It is claimed that within this proviso the Hamiltonian constraint is not a local equation satisfied at each spatial point but an equation integrated over a whole space. This observation is misleading  in one respect: when one implements a gauge fixing (which is the spirit of the projectability conditions) within the action itself, one may lose information on the constraint structure of the theory \cite{Pons:1995ss}, which must be restored by hand, and the information lost in this case is precisely the local structure of the Hamiltonian constraint. Of course one can always change the theory and consider that the projectability condition is a direct ingredient of the action principle, in which case the nonlocal Hamiltonian constraint appears, but this runs againts the spirit of writing a Lagrangian with no preconceptions as to whether its dynamical consequences could be. Sticking to an action principle with no further preconceptions, one can introduce the projectability conditions as a partial gauge fixing, as done above, but it does not have consequences regarding the locallity of the Hamiltonian constraint.

\vspace{1cm}

Before closing this section, let us make three relevant remarks concerning the previous two subsections, \ref{newimpl} and \ref {projec}.  First of all, both outcomes are similar to that encountered by \cite{Li:2009bg}, where it is argued that an infinite set of constraints appears in the theory, but whereas in \cite{Li:2009bg} the focus is in the potential term and $\lambda$ is kept to its GR value, our analysis is for  $\lambda\neq 1$. In spite of the existence of an infinite chain of constraints, particular configurations compatible with the full set of constraints exist, as shown in subsections \ref{ssphersymm} and \ref{timedep} below.

The second remark, concerning a possible way out of this infinite string of constraints, is the extreme, but consistent setting, of requiring that the variables $ \gamma_{ij}\, ,\pi^{ij}\,,N\,,N^i$ be only time dependent. 
This outcome is nothing but the dimensional reduction of  (\ref{thelag}) to a mechanical model with only one coordinate - the time. The reduction is anticipated to be consistent because it is made under the abelian group generated by the Killing vectors $\partial_i$. The only gauge freedom left is that of time reparametrizations. Though consistent, this extremely restrictive framework seems of very limited interest, at least for what regards its classical field theory content. Obviously, for $\beta \to 0$ we do not recover GR. 

Finally, the last remark concerns the issue of consistency. We expect that the addition of higher ordre terms to the minimal setting discussed here, (\ref{thelag}), does not change any of our conclusions. In fact, it will just add further complications to the constraint analysis, which will appear even more restrictive.

\section{Some well-known solutions satisfying the constraints}
\label{swsol}

To complete our findings we give some explicit examples of solutions. As such, they fulfill the condition (\ref{thetert}) and as a consequence any possible chain of constraints derived form its stabilization.
We choose two different set ups, subsections \ref{ssphersymm} and \ref{timedep}: the first one does not imply the {\sl projectability} condition whereas the last one does.

\subsection{ Black hole solutions}
\label{ssphersymm}

An obvious setting which guaranties (\ref{thetert}) is to require the vanishing of $\pi_{ij}$ which in turns implies, recalling the Hamiltonian constraint (\ref{hamconst}),
\beq
R\simeq 2\Lambda\,.
\label{reql}
\eeq
Notice that $\pi^{ij}\simeq 0$ directly implies 
$K^{ij}\simeq 0$, and hence the parameter $\lambda$ in (\ref{thelag}) and in the equations of motion (eom) (\ref{hordinamics1}), (\ref{hordinamics2}) plays no role at all, as it is directly seen in the Hamiltonian (\ref{realham}). 

\subsubsection{Schwarzswild black hole}
By direct inspection of (\ref{Kij}) we can conclude that $K^{ij}\simeq 0$ is satisfied by any stationary solution with vanishing shift $N^i$. In such case we need to explore the stabilization of the new constraints 
\begin{equation}
N^i\simeq 0\,,\quad \pi^{ij}\simeq 0\,.
\end{equation}
Preservation in time of the vanishing of the shift trivially determines the arbitrary functions $\lambda^i =0$ in the Dirac Hamiltonian, while the stabilization of $\pi^{ij}\simeq 0$ gives the new constraint
\begin{equation}
N R_{ij} -  (\nabla_i\nabla_j N - \gamma_{ij}\triangle N) \simeq 0\,.
\label{nepi}
\end{equation}
The trace of (\ref{nepi}) implies, using (\ref{reql}), the following constraint on $N$
\beq
\triangle N +N \Lambda\simeq 0\,,
\label{rn}
\eeq
{\sl which is no longer compatible with the projectability condition, as long as we keep $N\neq 0$ and $\Lambda\neq 0$}. Finally, inserting (\ref{rn}) in (\ref{nepi}), the new constraint is  
\beq
 N R_{ij} - \nabla_i\nabla_j N - \Lambda N \gamma_{ij}\simeq 0\,,
\label{rn2}
\eeq
which already implies (\ref{rn}), if one takes into account (\ref{reql}).
The stabilization of (\ref{rn2}) yields a partial determination of the arbitrary function 
$\lambda^0$ in the Dirac Hamiltonian but (\ref{rn2}) can also be seen as a partial determination of the lapse. 

\vspace{4mm}

Intepreting equation (\ref{rn2}) as an equation for the lapse, $R_{ij}$ must satisfy some projectability 
conditions. Let us find them. Computing the divergence of (\ref{rn2}) and using the result $\displaystyle\nabla_{\!j}\nabla_{\!i}\nabla^{\!j}N = \partial_{i}\triangle N + R_{ij} \nabla^{\!j}N $ one obtains the equation
\beq
N \nabla^{\!j}R_{ij} - \partial_{i}(\triangle N +N \Lambda)\simeq 0\,.
\eeq
The last piece vanishes using the constraint itself, since (\ref{reql}) and (\ref{rn2}) imply (\ref{rn}). Factorizing the lapse, which we assume is a non-vanishing function, we arrive at the integrability condition for (\ref{rn2}) 
\begin{equation}
\nabla^j R_{ij} \simeq 0\,.
\end{equation}
Using the contracted Bianchi identity, this condition is equivalent to 
$\nabla_{\!i} R\simeq 0$, which is satisfied because of (\ref{reql}). Thus the integrability condition is satisfied, showing the consistency of our procedure.

\vspace{4mm}

As an application of this setting, consider the static, spherically symmetric ansatz 
\beq
ds^2 = -A(r) dt^2 + B(r) dr^2 + r^2 d\Omega_2^2.
\label{sphersymm}
\eeq
This ansatz has been already considered in \cite{Lu:2009em} but under the detailed balance condition. 
Eq. (\ref{sphersymm}) is in fact a consistent reduction, which sets the shift vectors to zero as well as eliminates other metric components. The solution fulfilling the eom and the constraints (\ref{reql}) and (\ref{rn2}) boils down to just Schwarzschild, either de Sitter or Anti de Sitter 
\begin{equation}
A(r)=B^{-1}(r) = 1 - \frac{ r^2\Lambda}{3}+ \frac{a}{r}\,.
\end{equation}

\subsubsection{Kerr blak hole}
One may wonder whether any solution in GR admits a continuation to a solution in the modified theory (\ref{thelag}). It is clear that solutions to (\ref{thelag}), which must fulfill the constraint (\ref{thetert}), yield in the limit $\lambda\to 1$ solutions of GR. But precisely, because of the existence of the {\sl additional} constraint (\ref{thetert}), the opposite is not true.
Candidate GR solutions that, in principle, may fail with the fulfillment of (\ref{thetert}) are those with non-vanishing momenta $\pi^{ij}$. If the solution is stationary this means that the shift should be different form zero. A natural candidate is therefore a rotating black hole, and a possible choice inside GR is the Kerr black hole - we set $\Lambda=0$ for simplicity. This solution will have a continuation to a solution of (\ref{thelag}) as long as (\ref{thetert}) is satisfied. Let us check whether this is the case. Writting the metric in the dragging frame, the non-vanishing components of the $3$-metric are
\beq
g_{rr}= \frac{\rho^2}{\Delta}\,,\quad g_{\theta\theta}= \rho^2\,,\quad g_{\varphi\varphi}= \Big((r^2+\alpha^2) + \frac{r r_s \alpha^2 \sin^2\theta}{\rho^2}\Big)\sin^2\theta\,,
\label{3kerr}
\eeq
with $r_s$ the Schwarzschild radius and 
\begin{equation}
\alpha={J\over M}\,,\quad\rho^2=r^2+\alpha^2\cos^2\theta\,,\quad\Delta=r^2-r r_s +\alpha^2\,.
\end{equation}
Using (\ref{3kerr}) we can obtain the 3-dimensional shift and lapse fields 
\begin{eqnarray}
\label{shifK}
N^2 &=& - 1 +{r r_s\over \rho^2}-{r^2 r_s^2 \alpha^2 \sin^2\theta\over \rho^2(\rho^2 (r^2+\alpha^2)+r r_s \alpha^2 \sin^2\theta)} \,, \\ \nonumber
\vec{N}&=&-\left(0,0,{r r_s\over (r^2+\alpha^2)\rho^2 + r r_s \alpha^2 \sin^2\theta}\right)\,.
\end{eqnarray}
One can check from (\ref{3kerr}) and (\ref{shifK}) that $\pi^{ij}\neq 0$, but its trace vanishes, $\pi=0$. This means that (\ref{thetert}) is still satisfied. Since one can read from (\ref{hordinamics1},\ref{hordinamics2}) that the difference between the eom of GR and those of (\ref{thelag}) lies solely in terms that are proportional to $\pi$, we can immediately conclude that the Kerr solution (\ref{3kerr}) is also solution of (\ref{thelag}) {\sl for any value} of $\lambda$.

\subsubsection{The Galilean nature of time for $\lambda\neq 1$}
\label{gali}

In view of the previous two examples one may be tempted to erroneously conclude that any solution to GR with sufficient symmetry can always be deformed in the $\lambda$-parameter space to fulfill the new dynamics defined by (\ref{thelag}). In fact, for the theories defined with (\ref{thelag}) with $\lambda\neq 1$, the restriction to foliation preserving diffeomorphisms has the consequence that there is a preferred frame with Galilean time, thus restoring and absolute concept of simultaneity. The assertion that Schwarzschild, or Kerr, are solutions of (\ref{thelag}) must be qualified: they are solutions in the coordinate frames that only differ in a foliation preserving diffeomorphism from the expressions given in the last subsections for such solutions. But if these solutions, as solutions of GR, are presented in other frames, they are no longer solutions of (\ref{thelag}). Let us give the example of Schwarzschild in - ingoing - Painlev\'e-Gullstrand coordinates, 
\begin{equation}
ds^2= -dt^2 + (d r + \sqrt{1-A(r)} d t)^2 + r^2 d\Omega_2^2\,,
\label{painl}
\end{equation}
where the $3$-metric is simply flat, the lapse is trivial and the shift has the only component $N^r = \sqrt{1-A(r)}$. One can immediately check that $\triangle\pi\neq 0$. This result, in view of $N=1$, shows that the constraint (\ref{thetert}) is not fulfilled and therefore (\ref{painl}) is not a solution of (\ref{thelag}). 
This result above is not in contradiction with (\ref{sphersymm}) being a solution of (\ref{thelag}), because the two types of coordinatization used in (\ref{sphersymm}) and (\ref{painl}) are not connected by a foliation preserving diffeomorphism, and therefore the backgrounds described with these coordinatizations, though they lead to a single physics when we go to the GR limit, represent different physical settings for the dynamics given by (\ref{thelag}).
Later we shall complement these findings in terms of Rindler coordinates and examine its immediate consequences for thermodynamics, For the time being let us make the remark that one can associate a Hawking radiation in both systems of coordinates, and thus this emission is insensitive to the partial time diffeomorphism breaking.

\subsection{ Vacuum cosmologies, generalities}
\label{timedep}

A less restrictive non generic setting, still fulfiling (\ref{thetert}), is constructed by demanding 
\beq
\pi^{ij} = b(t)\sqrt{\gamma} \gamma ^{ij}\,, 
\label{pigam}
\eeq
in the gauge $N^i = 0\,$. We will obtain time dependent backgrounds that will lead to a number of {\sl interesting}  consequences for well known cosmologies. As a matter of notation, and since we are directly looking for solutions, we write henceforth ordinary equalities when dealing with the constraints.

 First of all notice that the geometry of the $3$-surfaces implies that $b$  in (\ref{pigam}) is a 
$3$-scalar, because $\pi^{ij}$ is a tensor density; thus the assertion that $b$ depends only on the time coordinate is a covariant statement with respect to $3$-diffeomorphisms. And it is also so with respect to the time foliation preserving diffeomorphisms.

Inserting (\ref{pigam}) in the eom (\ref{hordinamics1}) and solving for $\gamma$ one obtains
\beq
\gamma_{ij}({\bf x},t) = \exp\left({\frac{2 N}{1-3\lambda}\int_0^t b(\tau) d\tau}\right) \gamma_{ij}({\bf x},0)\,,
\label{exp}
\eeq
that is, the evolution of the $3$-metric in terms of expansion factor, $\exp({\frac{N}{1-3\lambda}\int_0^t b(\tau) d\tau})$, and the initial data. Notice that this expansion is {\sl rigid} in the sense that all points on the surfaces evolve with the same factor. 3-dimensional, distances between them are only affected by a global time dependent factor.

In the sequel we will obtain the dynamical evolution of $b(t)$ for different cases. Let us stress nevertheless that this is not the most {\sl efficient way} of constructing time-dependent metric spaces which are solutions of (\ref{thelag}) should one be interested in including matter.

Notice, first of all, that with the ansatz (\ref{pigam}) the Hamiltonian constraint (\ref{thetert}) becomes 
\beq  
R = \frac{3}{1-3\lambda} b^2(t) +2 \Lambda\,,
\label{rt}
\eeq 
which enforces $R$ to depend only on time. 
From the trace of (\ref{pigam}) we infer that the tertiary constraint (\ref{thetert}) is satisfied because 
$\nabla_{\!i}\pi=0$. Note that our ansatz complies with the assumptions of subsection \ref{newimpl}, including the projectability condition of the lapse, to be obtained below.

There are two ways to compute the time derivative of $\pi^{ij}$: either we use 
(\ref{hordinamics2}) or the rhs of the ansatz (\ref{pigam}). Using the former we obtain
\beq
\dot\pi^{ij} = \sqrt{\gamma}N \Big(-R^{ij}+ \frac{1}{2}R \gamma^{ij}
- \frac{1}{2(1-3\lambda)}b^2\,\,\gamma^{ij} - \Lambda\gamma^{ij}\Big)\,,
\label{dp}
\eeq
while the latter leads to
\beq
\dot\pi^{ij} = \frac{d}{d t}(b \sqrt{\gamma}\gamma^{ij}) = \Big(\dot b +\frac{N}{1-3\lambda} b^2 \Big) \sqrt{\gamma}\gamma^{ij}\,.
\label{dp2}\eeq
Equating both expressions, taking the trace, and using (\ref{rt}), we obtain our fundamental equation
\beq 
\dot b +\frac{N}{1-3\lambda} b^2 + \frac{1}{3} N \Lambda=0\,,
\label{db}
\eeq
We will explore the solutions to (\ref{db}) later on. Note that (\ref{db}) enforces the projectability condition on the lapse and it also indicates that $\dot\pi^{ij}$ is proportional to the metric $\gamma^{ij}$. Applying this result to (\ref{dp}), and using (\ref{rt}) and (\ref{db}), we obtain 
\beq 
R_{ij} =  (\frac{b^2}{1-3\lambda}+ \frac{2}{3} \Lambda)\gamma_{ij} = - \frac{\dot b}{N} \gamma_{ij}\,.
\label{ricc}
\eeq
We infer from equation (\ref{ricc}) that the surfaces of the foliation are Einstein spaces. The $3$-curvature, $\displaystyle R= -  \frac{3 \dot b}{N}$ has the opposite sign to $\dot b$. 

According to the signs of the cosmological constant $\Lambda$ and the coefficient $1-3\lambda$ we will have  four different cases  that we consider below. In all the following set ups, $N$ is gauge fixed to 
\beq
N= \frac{\vert 1-3\lambda\vert}{2}\,,
\label{theN}
\eeq
and $\alpha \ge 0$ is defined as
\beq
\frac{1}{\alpha^2} := \frac{1}{3} N \vert\Lambda\vert\,.
\label{theaplha}
\eeq
\subsubsection{ de Sitter-like space: $1-3\lambda<0,\ \Lambda >0$ }
\label{sub1}
This case contains GR - with positive cosmological cosntant - as a particular case and seems to be phenomenologically the most plausible scenario \cite{Dutta:2010jh,Lobo:2010hv}. 
Then the equation (\ref{db}) becomes 
\beq 
\dot b -\frac{1}{2} b^2 + \frac{2}{\alpha^2}=0\,.
\label{db1}
\eeq
Its solutions are separated in three regimes, accordingly with the initial conditions, given by the sign of $\displaystyle\frac{2}{\alpha}-\vert b\vert$\,.  

\begin{description} 
\item[ Closed slicing.]  
If the initial data fulfills the inequality
\begin{equation}
\vert b\vert< \frac{2}{\alpha}\,,
\end{equation}
the solution to (\ref{db1}) is given by 
\begin{equation}
b(t)=-{2\over  \alpha}   \tanh \left(\frac{t}{\alpha}\right)\,,
\end{equation}
up to a rigid time translation. For this solution $\dot b$ is negative. The simplest way to realize the $3$-dimensional Einstein space with positive curvature is the $3$-sphere 
which has $R_{ij} =  2 \gamma_{ij}$. 
Using (\ref{exp}), the $4$-metric becomes 
\beq
ds^2= -N^2 dt^2 + N\, \alpha^2 \cosh ^2\left(\frac{t}{\alpha}\right)d\Omega_3^2\,.
\label{thesol1} 
\eeq

\item[Open slicing.]  If the initial condition is such that 
\begin{equation}
\vert b\vert > \frac{2}{\alpha}\,, 
\end{equation}
the solution to (\ref{db1})  reads, up to a rigid time translation
 \begin{equation} 
 b(t)=-\frac{2 }{\alpha}\coth \left(\frac{t}{\alpha}\right)\,,
 \end{equation}
 and now $\dot b$ is positive. The simplest way to realize the $3$-dimensional Einstein space with negative curvature is the $3$-hyperboloid, 
which has $R_{ij} = - 2 \gamma_{ij}$. With a procedure similar to the previous case, using (\ref{exp}), the $4$-metric becomes 
\beq
ds^2= -N^2 dt^2 + N\, \alpha^2 \sinh ^2\left(\frac{t}{\alpha}\right)d H_3^2\,.
\label{thesol2} 
\eeq
This case admits the limit $\alpha\to\infty$, where (\ref{thesol2}) becomes
\beq
ds^2= -N^2 dt^2 + N\, t^2 d H_3^2\,,
\label{thesol4} 
\eeq
which, for $\lambda=1$, reduces to Milne universe, i.e. FRW solution in absence of matter.

\item[Flat slicing.]  As a last case we encounter the critical situation 
\begin{equation}
\vert b\vert= \frac{2}{\alpha}\,.
 \end{equation}
This case has the trivial constant solutions 
 \begin{equation} 
  b(t) = \pm \frac{2}{\alpha}\,.
  \end{equation}
 The $3$-curvature vanishes and we realize it with a spatial flat euclidean space. The $4$-metrics are
\beq
ds^2= -N^2 dt^2 + \exp\left(\pm \frac{2}{\alpha} t \right)(dx^2+dy^2+dz^2)\,.
\label{thesol3} 
\eeq

\end{description} 

\vspace{6mm}
The solutions (\ref{thesol1}),(\ref{thesol2}) are only solutions of GR for $\lambda=1$. Instead,  (\ref{thesol3}) is a solution of GR for any $\lambda$ if one rescales conveniently the cosmological constant. 
All the above solutions (\ref{thesol1}),(\ref{thesol2}),(\ref{thesol3}), boil down to the {\sl same} GR solution for $\lambda=1$: de Sitter space. For $\lambda\neq 1$ the gauge group is reduced to foliation preserving diffeomorphisms, and as a consquence the three cases, open, closed and flat slicing, are indeed three different vacuums. This is vacuum degeneracy. Taken this result at face value, if nature had initially broken the $4$-diffeomorphism group in this region of the $(\Lambda,1-3\lambda$) plane, it would also had to undergo an spontaneous symmetry breaking \cite{Sen:1971km} in order to conform with the present-day observationally allowed values of $\lambda$ and $\Lambda$. This mechanism of spontaneous symmetry breaking is also present in other models of Lorentz violation with an unconventional kinetic term \cite{ArkaniHamed:2005gu}. Althoug this picture is appealing, one should bear in mind that the analysis made here is {\sl classical}, without including matter,  and susceptible to quantum corrections.

\subsubsection{ Anti de Sitter-like space: $1-3\lambda<0,\ \Lambda < 0$}
\label{sub2}
The equation (\ref{db}) becomes
\beq 
\dot b -\frac{1}{2} b^2 - \frac{2}{\alpha^2}=0\,,
\label{db2}
\eeq
which has as solution
 \beq 
 b(t)=\frac{2 }{\alpha}\tan \left(\frac{t}{\alpha}\right)\,,
 \eeq 
up to a rigid time translation. Now $\dot b$ is positive and the $4$-metric becomes, 
\beq
ds^2= -N^2 dt^2 + N\, \alpha^2 \cos^2\left(\frac{t}{\alpha}\right)d H_3^2\,.
\label{thesol4} 
\eeq
For $\lambda=1$ this is just a partial covering of Anti de Sitter spacetime with coordinate singularities located at $\cos(\frac{t}{\alpha}) =0$.

\vspace{0.5cm}

Hitherto we have discussed cases that presume a {\sl mild} modification with respect to GR in the sense that the value of $\vert 1 - 3\lambda\vert$ in the $(\Lambda,1-3\lambda)$ plane lies in the same, lower half part where GR resides. In that respect the remaining two cases we will discuss can support deviations far from GR because they lie in the upper half plane.

It is worth noticing that if $b(t)$ solves (\ref{db}) for some signs of $(1-3\lambda)$ and $\Lambda$, the change $b(t)\to -b(t)$ produces a solution of (\ref{db}) with the opposite signs, whereas the expansion parameter in (\ref{exp}) does not change because of compensating factor signs. This fact allows us to obtain immeditately the new solutions below from the previous ones above. 

\subsubsection{$1-3\lambda>0,\ \Lambda > 0$}
\label{sub3}

This case is parallel to that in \ref{sub2} but with the change $b(t)\to -b(t)$. 
The $4$-metric becomes 
\beq
ds^2= -N^2 dt^2 + N\, \alpha^2 \cos^2\left(\frac{t}{\alpha}\right)d\Omega_3^2\,.
\label{thesol5} 
\eeq

\subsubsection{$1-3\lambda>0,\ \Lambda < 0$}
\label{sub3}

Now this case is just that of \ref{sub1} with the change $b(t)\to -b(t)$. All previous considerations hold and we end up with the $3$-metrics.
\bea
ds^2&=& -N^2 dt^2 + N\, \alpha^2 \cosh ^2\left(\frac{t}{\alpha}\right)d H_3^2\,.\\\nonumber
ds^2&= &-N^2 dt^2 + N\, \alpha^2 \sinh ^2\left(\frac{t}{\alpha}\right)d\Omega_3^2\,.\\\nonumber
ds^2&=& -N^2 dt^2 + \exp\left(\pm \frac{2}{\alpha} t \right)(dx^2+dy^2+dz^2)\,.
\label{thesol8} 
\eea


\hspace{1cm}

\begin{figure}[h]
  \centerline{\includegraphics[width=6in]{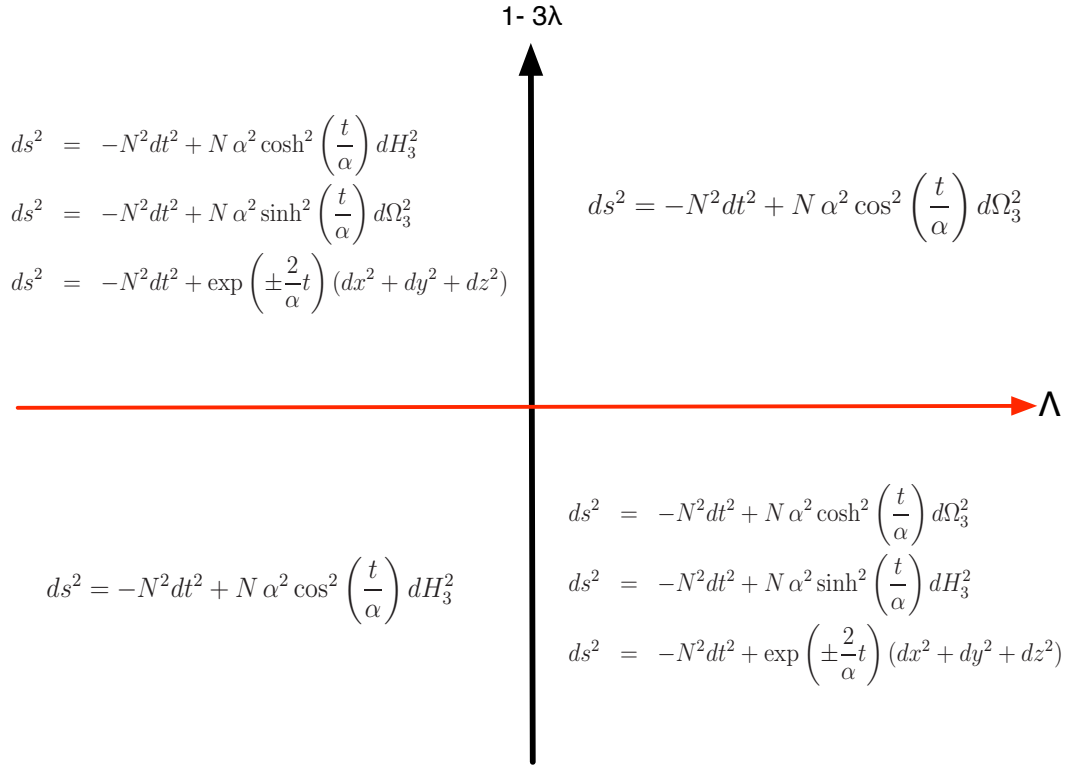}}
 \caption{ Pictorial representation of the vacuum solutions in the $(\Lambda,1-3\lambda)$ plane. The red 
axis represents the critical case $\displaystyle\lambda={1\over 3}$. $N$ stands for $\displaystyle N= \frac{\vert 1-3\lambda\vert}{2}$ and $\displaystyle\frac{1}{\alpha^2} := \frac{1}{3} N \vert\Lambda\vert$}
  \label{lambdasplane}
\end{figure}

The considerations in the above section concerning the space of vacuum solutions can be summarized in figure \ref{lambdasplane}.

Notice that while the transition between left$\leftrightarrow$right half planes as a function of $\Lambda$ is smooth this is not the case for the top$\leftrightarrow$botton half planes as a function of $1-3\lambda$.
Notice also the existence of a ``quasi-inversion'' symmetry by which we send $1-3\lambda\to -(1-3\lambda)$ and $\Lambda\to-\Lambda$ and the spherical and hyperbolical $3$-dimensional spaces are exchanged.

\subsubsection{The Galilean nature of time, revisited}
\label{gali2}

One of the disturbing outcomes of the model (\ref{thelag}) is Lorentz violation, expressed by the fact that only time foliation preserving diffeomorphism are permitted. As mentioned earlier \ref{gali} this is connected with the existence of a preferred time frame. Even if it has been advocated  in cosmology that there is no reason to refuse the existence of that frame and furthermore that there exists such a natural, preferred frame defined by the cosmic microwave background \cite{Kanno:2006ty} it is also true that, at least in the case of GR, tight phenomenological constraints rule out the existence of Lorentz violation operators of dimension $\le 6$ \cite{Mattingly:2005re}. Thus it seems that consistency with phenomenological results does not leave too much room for Lorentz violation, unless it comes in a very exotic manner. 
Anyhow we will explore some of the possible consequences of Lorentz violation at most fundamental level. Our starting point is a solution of GR, the Milne universe, (\ref{thesol4}) which covers the complementary wedge of Rindler spacetime in Minkowski spacetime. It is well known that, for $t >0$, the following change of coordinates
\beq
R=t \sinh\psi\,,\quad T= t \cosh\psi\,
\label{RT}
\eeq
makes this solution to be written as 
\begin{equation}
 ds^2= - dT^2 +dR^2 + R^2(d\theta^2 + \sin^2\theta d\phi^2)\,,
 \end{equation}
which is plainly just half of Minkowski spacetime, $T>0$.
We have gone with such detail, in these trivial matters, because of the points we want to make. Notice that for $\lambda\neq 1$, we can no longer practice arbitrary reparametrizations, but only foliation preserving diffeomorphisms. In particular, a change of coordinates of the type (\ref{RT}) is no longer permitted. This difference with GR is crucial in two essential points with a common origin:
\begin{description}
\item[Cosmological time.] 
In the GR case, the hyperboloid nature of the $3$-surfaces of the foliation in (\ref{thesol2}) was just an artifact of the choice of coordinates for de Sitter spacetime. But in theories with $\lambda \ne 1$ we are stuck with such a foliation with hyperboloids, which are equal time surfaces, and the parameter $t$ is already the physical  {\sl time}, except perhaps for a reparametrization involving only this time parameter itself. In other words,
what for GR was just a coordinate singularity for $t=0$ in (\ref{thesol2}) has become for $\lambda\neq 1$ theories a {\sl true background singularity}, a big bang for the solution with $t>0$ or a big crunch for that with $t<0$. 

In this spirit the three types of de Sitter-like spacetime foliations can be interpreted as: {\sl Closed slicing} (\ref{thesol1}) describes a bounce.  {\sl Open slicing} (\ref{thesol2}) describes a big bang or a big crunch with the singularity at finite time. The case of {\sl Flat slicing} (\ref{thesol3}) describes an infinite expansion or contraction with the singularity at infinity.

\item[Thermodynamics.]
Unlike Hawking radiation effect, Unruh effect is tantamount to Lorentz symmetry: in the absence of the latter the former does not exist \cite{Campo:2010fz}. The algebraic proof, in the framework of axiomatic field theories, states that the Minkowski vacuum restricted to a Rindler wedge is a thermal state with respect to the boost parameter \cite{Bisognano:1975ih}. It is obvious that in the absence of boost transformations of the Lorentzian type, we are not allowed anymore to define a thermal state i.e. one can not even 
find a horizon with a surface gravity from which to define the temperature through Tolman's law.

\end{description}

Notice that besides these discrepancies with GR, the structure of the time diffeomorphism breaking allows the preservation of the continuos self similarity property on the solutions of the modified dynamics.

\section{The critical case $\lambda= \frac{1}{3}$}
\label{crit}

As anticipated in section \ref{setup}, the critical case needs special attention because of the appearance of a new primary constraint, $\pi\simeq 0$. 
This means in particular that all solutions of (\ref{thelag}), including the GR case $\lambda=1$, that happen to satisfy $\pi= 0$, will be solutions for this critical case. And vice-versa: any solution of the critical case is a solution of  (\ref{thelag}) {\sl for any} $\lambda$.
The canonical Hamiltonian may be taken as the same as in GR, because the $\lambda$ dependent term in (\ref{realham}) vanishes for $\pi\simeq 0$. 
In fact this term is quadratic in the constraint. The canonical Hamiltonian has always the ambiguity of the addition of terms linear in the primary constraints. The choice of the GR form for (\ref{hamconst}) is the most convenient because of the closedness of the algebra of the Hamiltonian and momentum constraints in this case. This choice is legitimate because what is required to the canonical Hamiltonian is that $(F\!L^* p )\dot q - F\!L^*{\cal H} = {\cal L}$, where $F\!L^*$ is the pullback operation of the Legendre map $F\!L$ from the tangent bundle to the cotangent bundle, and this condition is satisfied in the critical case for {\sl any value} of $\lambda$ in (\ref{realham}). 

With this choice of the canonical Hamiltonian, the Dirac Hamiltonian is now
\beq
H_D = \int d{\bf x} (N^\mu {\cal H}^{\!{}^{(ADM)}}_\mu +\lambda^\mu P_\mu + \xi \pi)\,.
\label{dirhamcrit}
\eeq
Whereas the stabilization of the primary constraints $P_\mu\simeq 0$ yields the usual Hamiltonian and momentum constraints of GR, that of the new primary constraint $\pi\simeq 0$ produces the new secondary constraint 
\beq
N (R -3\Lambda) -\triangle N \simeq 0\,,
\label{newssc}
\eeq
that translates in a partial determination of the lapse. Finally, the stabilization of the Hamiltonian constraint yields a partial determination of the arbitrary function $\xi$,
\beq
\xi (R-3\Lambda) -\triangle \xi = 0\,.
\label{detxi}
\eeq
One can stabilize the constraint (\ref{newssc}), giving a partial determination of $\lambda^0$. Alternatively one could have eliminated the momenta $P_\mu$ from the formalism and take the lapse and shift as the arbitrary functions of the dynamics for the $3$-metric. 


\vskip.4cm
\noindent {\bf Acknowledgements}\\
This work 
has been partially supported by MCYT FPA 2007-66665, CIRIT GC 2005SGR-00564 ans Spanish Consolider-Ingenio 
2010 Programme CPAN (CSD2007-00042).
\appendix
\setcounter{equation}{0}
\newcounter{zahler}
\addtocounter{zahler}{1}
\renewcommand{\thesection}{\Alph{zahler}}
\renewcommand{\theequation}{\Alph{zahler}.\arabic{equation}}

\section{Appendix}

In this short appendix we display for the sake of completeness the eom for the minimal dynamics (\ref{thelag}). Let us we stress that dotted quantities in subsections \ref{newimpl} and \ref {projec} stand for the evolution under this minimally modified dynamics and they should not be confused with dotted quantities used in subsection \ref{lookfortert}, which meant evolution under the ADM dynamics in the gauge $N^i=0$. 

The eom for the dynamics (\ref{thelag}) are just a minimal modification of that of GR \cite{Wald:1984rg} 
\beq
\dot\gamma_{ij} =  \frac{2 N}{\sqrt{\gamma}}(\pi_{ij}+\frac{\lambda}{1-3\lambda} \pi \gamma_{ij}) + \nabla_i N_j + \nabla_j N_i\,,
\label{hordinamics1}
\eeq
\bea
\dot\pi^{ij} &=& \sqrt{\gamma}N \Big(-R^{ij}+ \frac{1}{2}R \gamma^{ij}+    
\frac{1}{2\gamma} \gamma^{ij}(\pi^{kl}\pi_{kl}+\frac{\lambda}{1-3\lambda}\pi^2) - 
\frac{2}{\gamma}(\pi^{ik}\gamma_{kl}\pi^{lj} + \frac{\lambda}{1-3\lambda} \pi \pi^{ij})- \Lambda \gamma^{ij}\Big) \nn\\
&+& \sqrt{\gamma}(\nabla^i\nabla^j N - \gamma^{ij}\nabla_k\nabla^k N) + \nabla_k(N^k \pi^{ij}) 
- \pi^{ik}\nabla_k N^j - \pi^{jk}\nabla_k N^i\,.
\label{hordinamics2}
\eea

\bibliographystyle{ssg}
\bibliography{hair}

\begingroup\raggedright\begin{thebibliography}{10}

\bibitem{Horava:2009uw}
  P.~Horava,
  ``Quantum Gravity at a Lifshitz Point,''
  Phys.\ Rev.\  D {\bf 79} (2009) 084008
  [arXiv:0901.3775 [hep-th]].

\bibitem{Charmousis:2009tc}
  C.~Charmousis, G.~Niz, A.~Padilla and P.~M.~Saffin,
  ``Strong coupling in Horava gravity,''
  JHEP {\bf 0908} (2009) 070
  [arXiv:0905.2579 [hep-th]].

\bibitem{Blas:2009yd}
  D.~Blas, O.~Pujolas and S.~Sibiryakov,
  ``On the Extra Mode and Inconsistency of Horava Gravity,''
  JHEP {\bf 0910} (2009) 029
  [arXiv:0906.3046 [hep-th]].

\bibitem{Iengo:2009ix}
  R.~Iengo, J.~G.~Russo and M.~Serone,
  ``Renormalization group in Lifshitz-type theories,''
  JHEP {\bf 0911} (2009) 020
  [arXiv:0906.3477 [hep-th]].

\bibitem{Henneaux:2009zb}
  M.~Henneaux, A.~Kleinschmidt and G.~L.~Gomez,
  ``A dynamical inconsistency of Horava gravity,''
  arXiv:0912.0399 [hep-th].

\bibitem{Calcagni:2009ar}
  G.~Calcagni,
  ``Cosmology of the Lifshitz universe,''
  JHEP {\bf 0909} (2009) 112
  [arXiv:0904.0829 [hep-th]].

\bibitem{Brandenberger:2009yt}
  R.~Brandenberger,
  ``Matter Bounce in Horava-Lifshitz Cosmology,''
  Phys.\ Rev.\  D {\bf 80} (2009) 043516
  [arXiv:0904.2835 [hep-th]].

\bibitem{Lu:2009em}
  H.~Lu, J.~Mei and C.~N.~Pope,
  ``Solutions to Horava Gravity,''
  Phys.\ Rev.\ Lett.\  {\bf 103} (2009) 091301
  [arXiv:0904.1595 [hep-th]].

\bibitem{Arnowitt:1962hi}
  R.~L.~Arnowitt, S.~Deser and C.~W.~Misner,
  ``The dynamics of general relativity,'' in 
{\sl Gravitation: an introduction to current research}, Louis Witten ed. (Wilew 1962), 1962. 
  arXiv:gr-qc/0405109.


\bibitem{reyes}
   R.~Reyes, {\it et al.},
  ``Confirmation of general relativity on large scales from weak lensing 
     and galaxy velocities,''
  Nature {\bf 464} (2010) 256.


\bibitem{Sotiriou:2009bx}
  T.~P.~Sotiriou, M.~Visser and S.~Weinfurtner,
  ``Phenomenologically viable Lorentz-violating quantum gravity,''
  Phys.\ Rev.\ Lett.\  {\bf 102} (2009) 251601
  [arXiv:0904.4464 [hep-th]].

\bibitem{Blas:2009qj}
  D.~Blas, O.~Pujolas and S.~Sibiryakov,
  ``A healthy extension of Horava gravity,''
  arXiv:0909.3525 [hep-th].

\bibitem{Mukohyama:2009tp}
  S.~Mukohyama,
  ``Caustic avoidance in Horava-Lifshitz gravity,''
  JCAP {\bf 0909} (2009) 005
  [arXiv:0906.5069 [hep-th]].

\bibitem{Dutta:2010jh}
  S.~Dutta and E.~N.~Saridakis,
  ``Overall observational constraints on the running parameter $\lambda$ of
  Ho\v{r}ava-Lifshitz gravity,''
  arXiv:1002.3373 [hep-th].

\bibitem{Wald:1984rg}
{\it See for instance Appendix E in}  R.~M.~Wald,
  ``General Relativity,''
{\it  Chicago Univ. Pr. ( 1984) 491p}

\bibitem{Kocharyan:2009te}
  A.~A.~Kocharyan,
  ``Is nonrelativistic gravity possible?,''
  Phys.\ Rev.\  D {\bf 80} (2009) 024026
  [arXiv:0905.4204 [hep-th]].

\bibitem{Pons:1995ss}
  J.~M.~Pons,
  ``Plugging the gauge fixing into the Lagrangian,''
  Int.\ J.\ Mod.\ Phys.\  A {\bf 11} (1996) 975
  [arXiv:hep-th/9510044].

\bibitem{Li:2009bg}
  M.~Li and Y.~Pang,
  ``A Trouble with Ho\v{r}ava-Lifshitz Gravity,''
  JHEP {\bf 0908} (2009) 015
  [arXiv:0905.2751 [hep-th]].

\bibitem{Lobo:2010hv}
  F.~S.~N.~Lobo, T.~Harko and Z.~Kovacs,
  ``Solar System tests of Ho\v{r}ava-Lifshitz black holes,''
  arXiv:1001.3517 [gr-qc].

\bibitem{Kanno:2006ty}
  S.~Kanno and J.~Soda,
  ``Lorentz violating inflation,''
  Phys.\ Rev.\  D {\bf 74} (2006) 063505
  [arXiv:hep-th/0604192].

\bibitem{Mattingly:2005re}
  D.~Mattingly,
  ``Modern tests of Lorentz invariance,''
  Living Rev.\ Rel.\  {\bf 8} (2005) 5
  [arXiv:gr-qc/0502097].

\bibitem{Sen:1971km}
  R.~N.~Sen and C.~Weil,
  ``Degenerate vacuum formalism for spontaneous symmetry breakdown,''
  J.\ Phys.\ A  {\bf 4} (1971) 632.

\bibitem{ArkaniHamed:2005gu}
 {\sl See for instance}  N.~Arkani-Hamed, H.~C.~Cheng, M.~A.~Luty, S.~Mukohyama and T.~Wiseman,
  ``Dynamics of Gravity in a Higgs Phase,''
  JHEP {\bf 0701} (2007) 036
  [arXiv:hep-ph/0507120].

\bibitem{Campo:2010fz}
  D.~Campo and N.~Obadia,
  ``Why does the Unruh effect rely on Lorentz invariance, while Hawking
  radiation does not ?,''
  arXiv:1003.0112 [gr-qc].

\bibitem{Bisognano:1975ih}
  J.~J.~Bisognano and E.~H.~Wichmann,
  ``On The Duality Condition For A Hermitian Scalar Field,''
  J.\ Math.\ Phys.\  {\bf 16} (1975) 985.
  ``On The Duality Condition For Quantum Fields,''
  J.\ Math.\ Phys.\  {\bf 17} (1976) 303.





\end{thebibliography}\endgroup

\end{document}